# Inferring the Sign of Kinase-Substrate Interactions by Combining Quantitative Phosphoproteomics with a Literature-Based Mammalian Kinome Network

Marylens Hernandez[1], Alexander Lachmann[1], Shan Zhao[1], Kunhong Xiao[2], Avi Ma'ayan[1*]

[1]Department of Pharmacology and Systems Therapeutics, Mount Sinai School of Medicine, New York, NY 10029, USA
[2]Department of Biochemistry, Duke University Medical Center, Durham, NC 27710, USA

**Protein phosphorylation is a reversible post-translational modification commonly used by cell signaling networks to transmit information about the extracellular environment into intracellular organelles for the regulation of the activity and sorting of proteins within the cell. For this study we reconstructed a literature-based mammalian kinase-substrate network from several online resources. The interactions within this directed graph network connect kinases to their substrates, through specific phosphosites including kinase-kinase regulatory interactions. However, the "signs" of links, activation or inhibition of the substrate upon phosphorylation, within this network are mostly unknown. Here we show how we can infer the "signs" indirectly using data from quantitative phosphoproteomics experiments applied to mammalian cells combined with the literature-based kinase-substrate network. Our inference method was able to predict the sign for 321 links and 153 phosphosites on 120 kinases, resulting in signed and directed subnetwork of mammalian kinase-kinase interactions. Such an approach can rapidly advance the reconstruction of cell signaling pathways and networks regulating mammalian cells.**

## I. Introduction

PROTEIN phosphorylation causes the addition of a phosphate group onto Serine, Threonine, or Tyrosine amino-acid residues. The addition of the phosphate usually results in a change of the substrate's activity leading to translocation, degradation, changes in enzymatic activity, or binding to other biomolecules such as other proteins, DNA or RNA. There are 518 known protein kinases [1] and 147 protein phosphatases [2] encoded in the human genome and it is approximated that 40% of all mammalian proteins are phosphorylated at some point in time in different cell types and at different cell states [1]. Recent advances in mass spectrometry (MS)-based phosphoproteomics have offered great opportunities for identification of protein phosphorylation sites on a proteome-wide scale. In addition, MS combined with stable isotope labeling technologies (i.e. quantitative phosphoproteomics) such as Stable Isotope Labeling of Amino acid in Cell (SILAC) and Isobaric Tag for Relative and Absolute Quantitation (iTRAQ) has emerged as a powerful tool to quantitatively assess dynamic changes of the identified phosphorylation in a high throughput manner [3, 4]. However, such data does not provide the kinases responsible for the phosphorylations. Such relationships are often identified experimentally using low throughput techniques such as radioactive labeling and affinity chromatography, or computational methods. Computational approaches that are used to predict the kinases most likely responsible for phosphorylations utilize consensus substrate amino-acid sequence motifs and other context dependent data. Several algorithms have been developed to accomplish this task [5, 6]. For example, NetworKIN [6, 7] implements an algorithm that combines several background knowledge "pieces-of-evidence" to predict the most probable kinase that is responsible for phosphorylating an identified phosphosite.

*Corresponding author: A. M. Author (e-mail: avi.maayan@mssm.edu).

Databases that integrate the results from phosphoproteomics experiments are emerging. Two leading examples are PhosphoSite [8] and Phospho.ELM [9]. Additionally, databases that record associated kinases with their substrates also grow rapidly. For a prior study, we constructed a web-based tool called Kinase Enrichment Analysis (KEA) [10]. For KEA we assembled most of the currently and publicly available experimentally determined kinase-substrate interactions from several kinase-substrate databases. By having a large background knowledge dataset of kinase-substrate interactions, we can begin to identify patterns of connectivity which unmask how groups of kinases regulate different aspects of cell behavior. Additionally, since many kinases are themselves regulated by protein phosphorylation, we can start assembling the regulatory network of kinase-kinase interactions to examine how kinases regulate each other to form functional signaling modules through phosphorylation cascades, feed-forward, and feedback loops. It is well-known that regulation of kinases through phosphorylation results in a complex web of regulatory relations. For example, it was experimentally demonstrated that a network of kinases function during filamentous growth in yeast [11]. Computational analyses of the yeast kinome identified that kinases form a scale-free network [12] where kinases are clustered into functional groups. Since mammalian cells have more genes that encode kinases as compared with yeast, it is expected that the mammalian kinome network is more complex than in yeast. In this study we aimed to reconstruct an initial version of the mammalian kinome network and then use the network's topology in combination with data from quantitative phosphoproteomics to infer the signs of links connecting kinases.

## II. Results

*Construction of a mammalian kinase-substrate network*

Using information available in the public domain we reconstructed an in-silico network using known kinase-substrate interactions. We only considered interactions that report the exact phosphorylation site (phosphorylated amino-acid on the substrate). The data sources used are HPRD [13], PhosphoSite [8], phospho.ELM [9], NetworKIN [6], and Kinexus (www.kinexus.ca). Data from HPRD contributed 4578 interactions from 1875 publications; Phosphosite contributed 6196 interactions from 2688 publications; phospho.ELM 2703 interactions from 1848 publications; Kinexus 1957 from 647 publications, and NetworKIN 5852 interactions from one paper. To integrate the data from these different sources, human, mouse and rat IDs where merged using NCBI homologene to match mammalian genes to their human ortholog. All data from these sources were organized into a five column flat file format containing the following information: the kinase, the substrate, the phosphosite, the effect of the phosphorylation on the substrate if known (activation/inhibition) and the PubMed ID linking to the publication that identified the phosphorylation interaction. In total, the consolidated dataset contains 14,374 interactions from 3469 publications involving 436 kinases and phosphatases.

Since kinases and phosphatases regulate each others' activity through phosphorylation and dephosphorylation, and since we are interested in understanding the structure and function of cell signaling networks in mammalian cells, we used the above dataset to extract a subnetwork involving only kinases. This mammalian kinome subnetwork extracted from the above dataset consists of 356 nodes connected through 1380 interactions extracted from 1072 papers. Some of the interactions, namely those reported by Kinexus, have signs associated with the links. 114 interactions are marked as activation and 85 as inhibitions whereas 1181 interactions do not have a sign associated with them. The average link per node in the subnetwork is 7.15 whereas the connectivity distribution fits a power-law. The subnetwork has a giant connected component made of size 320 nodes with a characteristic path length of 3.175. The subnetwork is highly dense with a high clustering coefficient of 0.566.

*Inferring the signs of kinase-kinase regulatory interactions*

Although the kinome subnetwork is represented as a directed graph, the signs of the interactions, namely activation or inhibition effects are mostly unknown. In order to address this issue, we devised an inference algorithm that can be used to infer the effect, activation or inhibition, of phosphosites on kinases and the signs that connect kinases. For this we combined the information about the connectivity of the kinome network with data collected from quantitative phosphoproteomics. We reasoned that if the majority of the substrates of a kinase mostly increase in phosphorylated phosphosites under some experimental condition and a phosphosite on the kinase also increases in level under the same experimental conditions, then the phosphsite on the kinase should be an activation site. Similarly if the majority of the substrates of a kinase are less phosphorylated under some experimental condition, and the phosphosite on the kinase also decreases in level under the same experimental condition, then the phosphsite on the kinase should also be an activation site. On the other hand if the site on the kinase decreases whereas the substrates increase, then the site is likely to be inhibitory; or if the site on the kinase is increasingly phosphorylated whereas the substrates decrease, then the site is likely to be inhibitory. Such logic is depicted in Fig. 1. This logic disregards opposing effects and competition between kinases and phosphatases phosphorylating or dephosphorylating the same sites. Hence, it is a simplification. Regardless, we believe that the method is valid for making reasonable predictions.

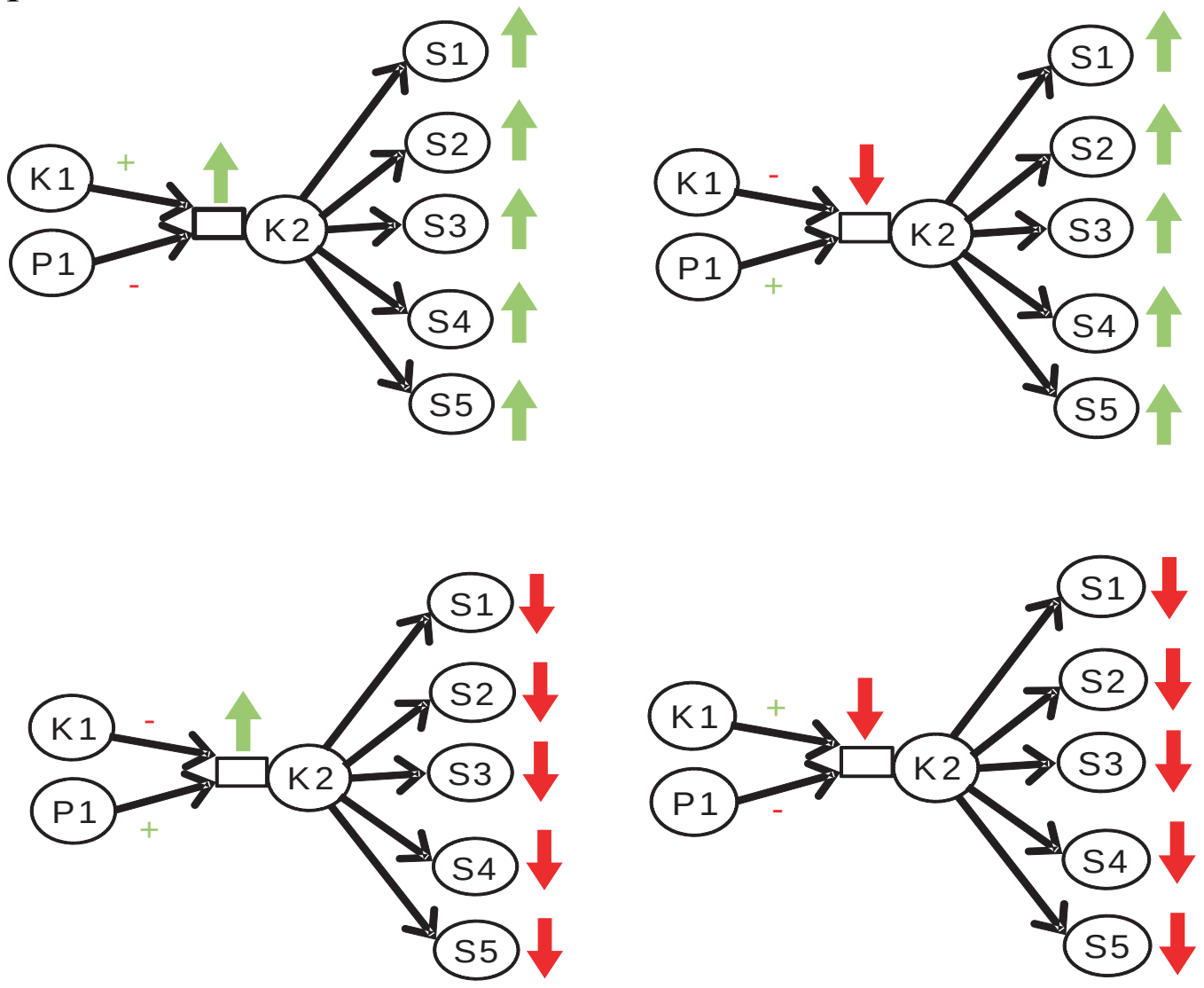

**Fig. 1 Illustration of the algorithm used to predict the sign of regulatory links that connect kinases and phosphatases by merging data from a literature-based kinase-substrate network with SILAC phosphoproteomics publications. K- kinase; P- phosphatase; S- substrate.**

To describe the inference method more formally we can let '$M_{mxn}$' be the connectivity matrix connecting "m" kinases and "n" phosphosites, such that $M_{ij}$ = 1 if kinase "i" is known to phosphorylate phosphosite "j", $M_{ij}$ = 0 otherwise. Let '$X_n$' be the vector that describes the behavior of all phosphosites in a particular phosphoproteomics experiment, where Xj = {0, 1, -1}, such that Xj = 0 if during the experiment the phosphorylation level of phosphosite 'j' did not change or wasn't determined, Xj = 1 if the phosphosite 'j' was increasingly phosphorylated, or Xj = -1 if the phosphorylation level of phosphosite 'j' was decreased. Having the connectivity matrix M and the vector X, and since there usually are multiple substrates for a specific kinase, the most common behavior of all substrates for a specific kinase, based on a specific experiment, can be calculated for each kinase by: $T_m = \text{sign}(M_{mxn}X_n)$. Note that because we are just interested in whether most phosphosite-substrates for a specific kinase were increased or decreased overall, we take the "sign" of the inner product. Here T is the resulting vector of size "m", such that $T_i = \{1, -1, 0\}$, $T_i = 1$ means that most phosphosite-substrates for kinase 'i' were increased, $T_i = -1$ means that most phosphosite-substrates for kinase 'i' were decreased, and $T_i = 0$ means that there is no relevant information for those phosphosite-substrates of kinase 'i' in the particular X vector experiment. Once we have computed T, the next step is to infer regulation based on the behavior of sites on those kinases. In order to do this we can define an "association matrix" $P_{nxm}$, such that $P_{ji} = 1$ if phosphosite j is on kinase i, $P_{ij} = 0$ otherwise. P associates kinases with the phosphosites on them. Then,

$$Q_n = T_m \,.\, P_{nxm} \,.\, X_n \qquad (1)$$

Where $[P_{nxm} \,.\, X_n]$ describes the behavior of each phosphosite 'j' on kinase 'i' in the experiment, and Q is the 'inference regulation vector' per phosphosite, such that $Q_j = 1$ means the effect of phosphosite 'j' is positive, $Q_j = -1$ means the effect of phosphosite 'j' is negative, $Q_j=0$ means the effect of phosphosite 'j' is unknown. Finally, taking the connectivity matrix into account, we can infer the sign of the direct links in the network, which are going to have the same sign of the corresponding phosphosite:

$$R_{mxn} = [Q_n \,.\, M_{mxn}^{T}]^{T} \qquad (2)$$

$R_{mxn}$ will be the 'inference regulation matrix' for 'm' kinases and 'n' phosphosites on kinases, where $R_{ij} = 1$ means kinase 'i' activates kinase 'j' through phosphosite 'j', $R_{ij} = -1$ means kinase 'i' inhibits kinase 'j' through phosphosite 'j' and $R_{ij} = 0$ means that the regulation is unknown. The final and complete formula is:

$$R_{mn} = [[[\mathrm{sign}(M_{mn}X_n)]_m \,.\, P_{nm} \,.\, X_n]_n \,.\, M_{mn}^{T}]_{nm}^{T} \qquad (3)$$

The same method can be applied to infer signs for phosphatases but the inference rules will be opposite.

*Application of the algorithm by using quantitative phosphoproteomic data combined with the literature-based kinome network*

To implement our inference method we first collected data from 12 phosphoproteomics publications reporting 23,283 phosphosites from 37 different separate experimental conditions, whereas 1342 phosphosites detected in those experiments were also present in the literature-based kinome network. A breakdown of the counts of phosphosites that increased or decreased in all SILAC phosphoproteomics experiments, and the fraction of phosphosites detected on kinases, are provided (Fig. 2). Feeding such data into our inference algorithm, we were able to predict the sign/effect for 153 phosphosites. Out of these 153 sites, 137 sites did not have a sign/effect previously associated with them. The remaining 16 sites are associated with 40 links where 30 of them were confirmed based on previously assigned signs, whereas 10 interactions were inconsistent with their previously assigned sign. 77 sites passed a Binomial proportion test ($p < .05$) with an underlying even distribution for detecting activation or inhibition sign, and only 25 sites passed the test if the underlying probability for the Binomial test is taken from the total proportion of predicted positive vs. negative signs. Finally, we constructed a signed and directed network made of the kinases that were identified to be regulating each other through the predicted signed links (Fig. 3).

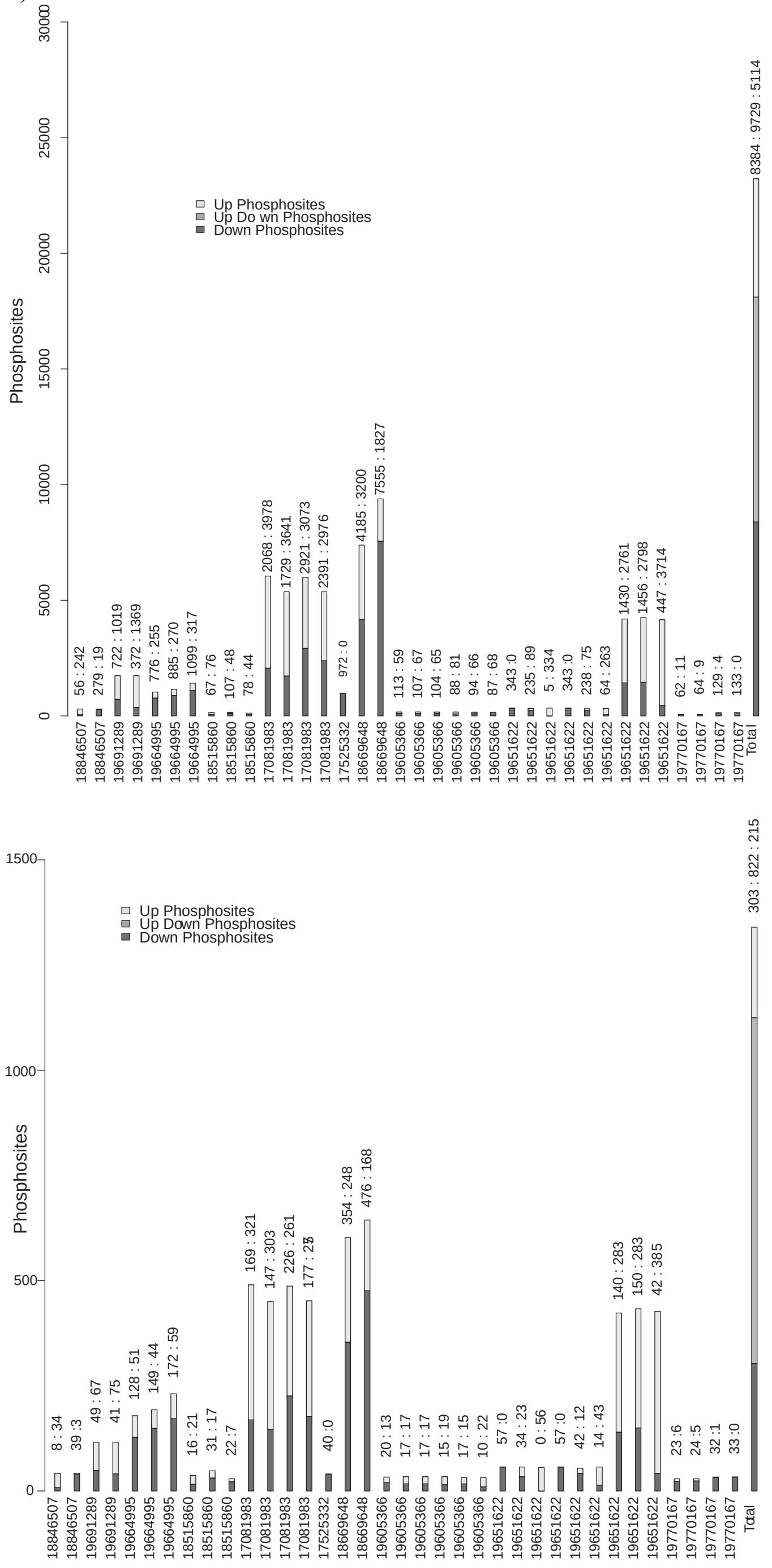

**Fig. 2 Breakdown of identified phophosites reported in different SILAC phosphoproteomics publications. A) All identified sites that displayed either increase or decrease in phosphorylation levels under some experimental condition. B) Counts of sites identified by the SILAC phosphoproteomics that also have a known upstream kinase from the literature-based kinase-substrate network.**

The network diagram only shows the large connected component of the predicted kinase-kinase regulatory interactions. The results recover nicely the MAPK cascade and place components in the right hierarchical order. Other previously known regulatory relations are confirmed.

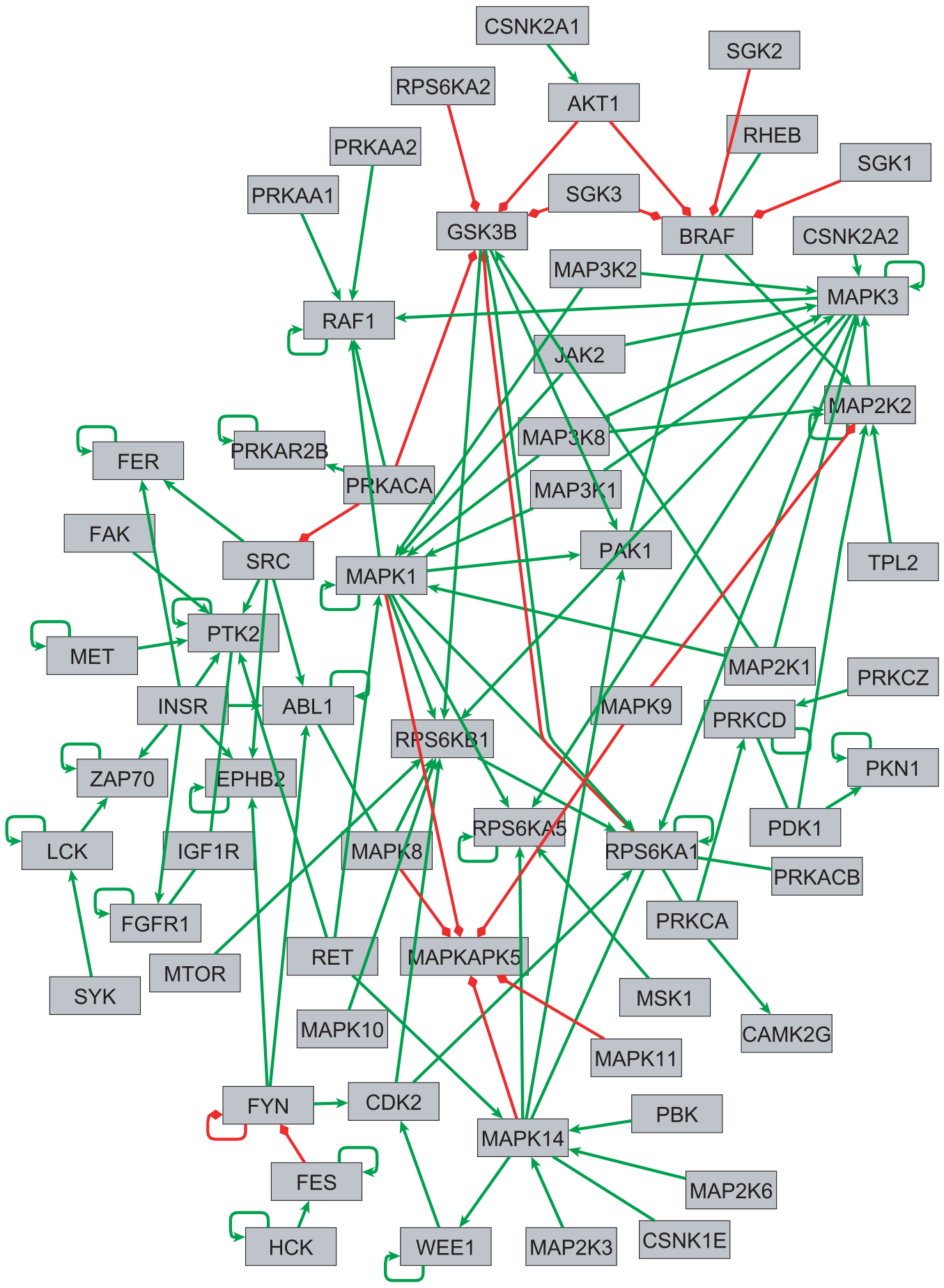

**Fig. 3 The largest connected component of the kinase-kinase network created from all inferred signs/effects. Nodes represent kinases; green arrows represent activations; red diamond heads represent inhibitions.**

For example, it was shown that GSK3β is inactivated in response to PI3K signaling, as a result of AKT1-mediated phosphorylation [14]. Additionally, the negative regulation of BRAF by AKT1 is supported by experimental evidence [15]. Hence, for most of the automatically inferred regulatory interactions the inference method appears valid regardless of whether the inferred interaction is positive or negative.

## III. CONCLUSIONS

In this study we show how, by combining data from quantitative phosphoproteomics experiments with literature-based kinase-substrate network, we can infer the signs/effects of links connecting kinases and phosphatases. Such knowledge extraction is critical for understanding signaling pathways and computationally modeling cell signaling networks. Our inference method makes some simplifying assumptions that should be considered. In most situations substrates can be phosphorylated or dephosphorylated by multiple kinases and phosphatases that can be activated or inhibited in different experimental conditions in different combinations. The inference method isolates kinase-substrate interactions from the global network effects. Such simplified assumption makes the calculation relatively straight forward. However, it can be substituted by a more complex inference algorithm that considers more complicated dependencies. In addition, as seen by the low coverage of known sites with known kinases, as compared to all known sites (Fig. 2), it is possible that the inference conclusions are highly inaccurate due to lack of available data. As more data become available, the accuracy and the confidence of the results based on statistical tests are expected to improve. Additionally, readers should be aware of the fact that the prior-knowledge kinase-substrate network is mainly derived from low-throughput studies that are notorious for errors since it is relatively easy to experimentally demonstrate that a kinase can phosphorylate specific sites on substrate proteins in-vitro. However, whether such phosphorylations are actually carried out in-vivo is always questionable. An additional way that we could implement to validate whether the sign inference method is working is to look at the position of the phosphosite in the kinase domain (e.g. activation loop phosphorylation is activating, while other sites can be inhibiting). This positional/structural attribute of the different sites could be analyzed to confirm network-based predictions. One of the interesting outcomes of our analysis is the fact that we found that phosphorylations were more commonly causing activation of kinases compared with inhibitions. This is interesting since the experimental reports do not show a bias for increases or decreases in phosphorylation on sites in general. Such observation is consistent with the hypothesis that phosphorylations in signaling pathways are more commonly activating downstream components, where as phosphatases are less specific, less regulated and more commonly used to shut off signaling [16, 17]. However, regardless of our initial observation, we feel that such hypothesis is still open for further experimental verification.

## ACKNOWLEDGMENT

This research was supported by NIH Grant No. 1P50GM071558-01A27398 (SBCNY) and start-up fund from Mount Sinai School of Medicine to A.M.